\documentclass[a4paper]{article}
\usepackage[german, english]{babel}
\usepackage{a4wide}
\usepackage{amsmath}
\usepackage{amssymb}
\usepackage{ifthen}
\usepackage{epsfig}
\newcounter{fig}   

\newcommand{\rd}{{\rm d}}
\newcommand{\vphi}{\varphi}

\newcommand{\sqt}{\sqrt{3}}

\begin{document}

\title{\bf Particle-like Platonic Solutions in Scalar Gravity}
\vspace{1.5truecm}
\author{
{\bf Burkhard Kleihaus, Jutta Kunz and Kari Myklevoll}\\
Institut f\"ur  Physik, Universit\"at Oldenburg, Postfach 2503\\
D-26111 Oldenburg, Germany}

\vspace{1.5truecm}

\date{\today}

\maketitle
\vspace{1.0truecm}

\begin{abstract}
We construct globally regular gravitating solutions,
which possess only discrete symmetries.
These solutions of Yang-Mills-dilaton theory
may be viewed as exact (numerical) solutions of scalar gravity,
by considering the dilaton as a kind of scalar graviton,
or as approximate solutions of Einstein-Yang-Mills theory.
We focus on platonic solutions with cubic symmetry,
related to a rational map of degree $N=4$.
We present the first two solutions of the cubic $N=4$ sequence,
and expect this sequence to converge to an extremal
Reissner-Nordstr\"om solution with magnetic charge $P=4$.
\end{abstract}

\section{Introduction}

Localized `particle-like' solutions play an important role in non-linear
field theories.
In flat space,
solutions of particular interest are Skyrmions 
(in the strong interaction) \cite{skyrme}, 
electroweak sphalerons \cite{klink},
or magnetic monopoles \cite{mono}. 

In the case of pure Yang-Mills theory,
or of pure Einstein gravity, in contrast,
there are no regular `particle-like' solutions \cite{CD}.
But when Yang-Mills theory and Einstein gravity are combined,
static, gobally regular `particle-like' solutions appear \cite{bm}.

The spatial symmetries of static classical solutions can be specified
by means of certain rational maps of degree $N$ \cite{ratmap}.
Maps of degree $N=1$ and $N=2$ lead to
classical solutions with spherical symmetry and axial symmetry,
respectively,
while for maps with degree $N \ge 3$, 
solutions with only discrete symmetries appear.
Identifying these with the symmetries of platonic solids or crystals,
we refer to them as platonic solutions.

In flat space platonic solutions, such as
platonic Skyrmions, platonic monopoles and platonic sphalerons,
are known \cite{skyrpla,monopla,sphapla}.
But does gravity also allow for
the existence of regular solutions with only discrete symmetries,
and does it even allow for platonic solutions,
which do not possess a flat space limit?

To answer these questions, we would like to construct
platonic generalizations of the famous static, spherically symmetric
Bartnik-McKinnon (BM) solutions \cite{bm},
representing globally regular solutions of Einstein-Yang-Mills (EYM) theory
with action
\begin{equation}
S=\int \left ( \frac{R}{4} 
 - \frac{1}{2}  {\rm Tr} (F_{\mu\nu} F^{\mu\nu})
 \right ) \sqrt{-g} \rd^4x 
\ , \label{action} \end{equation}
where $R$ denotes the scalar curvature,
$F_{\mu \nu} =
\partial_\mu A_\nu -\partial_\nu A_\mu + i \left[A_\mu , A_\nu \right] $
the SU(2) field strength tensor,
$A_\mu =  A_\mu^a \tau_a/2$ the gauge potential,
and Newton's constant $G=1/4 \pi$.

For static solutions, the 4-dimensional metric ($x^\mu=t,x^i$)
can be reduced according to 
\begin{equation}
 \rd s^2 = - e^{2 \phi} \rd t^2 + e^{-2 \phi} \rd s_3^2 \ , \ \ \
   \rd s_3^2 = h_{ij} \rd x^i \rd x^j 
\ , \label{metric1} \end{equation}
yielding the 3-dimensional action
\begin{equation}
 S = \int
\left\{\frac{R_3}{4} - \frac{1}{2} \partial_i \phi \partial^i \phi
 - \frac{1}{2} e^{2 \phi}\, {\rm Tr} (F_{ij} F^{ij}) \right\} \sqrt{h} \rd^3 x 
\ , \label{action3} 
\end{equation}
with $A_\mu \rd x^\mu = A_i \rd x^i$.

While we expect a 3-dimensional metric $h_{ij}=h_i(\vec x\,) \delta_{ij}$
with distinct metric functions $h_i$ for exact platonic solutions,
we here consider only approximate platonic EYM solutions,
obtained with a euclidian 3-dimensional metric.
For static configurations, in this approximation the action then agrees
with the action of Yang-Mills-dilaton (YMD) theory,
\begin{equation}
 S = \int \left\{
 -\frac{1}{2} \partial_\mu \phi \partial^\mu \phi
 - \frac{1}{2} e^{2 \phi}\, {\rm Tr} (F_{\mu\nu} F^{\mu\nu})\right\} \rd^4 x
\ , \label{actiond}
\end{equation}
where $\phi$ denotes the dilaton field,
and the solutions therefore represent platonic solutions of YMD theory.
Indeed, the dilaton may be considered as a kind of scalar graviton,
possessing also a universal coupling to matter.

Consequently, solutions of EYM and YMD theory are expected to
possess very similar properties. This is indeed the case
for spherically symmetric and axially symmetric solutions 
\cite{bm,ymd,kk,kk-dil}. 
The study of platonic YMD solutions is therefore expected to yield important
information about platonic EYM solutions, without having to solve
the highly complicated set of coupled Einstein-Yang-Mills equations.

Spherically and axially symmetric EYM and YMD solutions
form sequences, labelled by two discrete numbers, representing 
the degree of the map $N$, and the
node number $k$ of the gauge field functions \cite{bm,ymd,kk,kk-dil}.
We expect analogous sequences of platonic solutions,
i.e., for a given map of degree $N$, there should be 
a fundamental platonic solution 
and an infinite sequence of
excited platonic solutions with the same symmetries.
First indications of the existence of these solutions have been seen recently 
in Yang-Mills-Higgs-dilaton (YMHD) theory \cite{kkm-dil}

Here we construct such platonic YMD solutions.
We consider the Ans\"atze for these solutions in section 2,
we discuss their properties and present the numerical results in section 3,
and we give the conclusions and an outlook in section 4.

\section{Ans\"atze for Platonic Solutions}

To obtain YMD solutions with discrete symmetry
we make use of rational maps,
i.e.~holomorphic functions from $S^2\mapsto S^2$ \cite{ratmap}.
Treating each $S^2$ as a Riemann sphere, the first having coordinate 
$\xi$,
a rational map of degree $N$ is a function $R:S^2\mapsto S^2$ where
\begin{equation}
R(\xi)=\frac{p(\xi)}{q(\xi)} 
\ , \label{rat} \end{equation}
and $p$ and $q$ are polynomials of degree at most $N$, where at least
one of $p$ and $q$ must have degree precisely $N$, and $p$ and $q$
must have no common factors \cite{ratmap}.

We recall that via stereographic projection, the complex coordinate $\xi$
on a sphere can be identified with conventional polar coordinates by
$\xi=\tan(\theta/2)e^{i\varphi}$ \cite{ratmap}.
Thus the point $\xi$ corresponds to the unit vector
\begin{equation}
\vec {n}_\xi=\frac{1}{1+\vert \xi \vert^2}
(2\Re(\xi), 2\Im(\xi),1-\vert \xi \vert^2)
\ , \label{unit1} \end{equation}
and the value of the rational map $R(\xi)$ 
is associated with the unit vector
\begin{equation}
\vec {n}_R=\frac{1}{1+\vert R \vert^2}
(2\Re(R), 2\Im(R),1-\vert R\vert^2) \ .
\label{unit2}
\end{equation}

We here consider platonic YMD solutions obtained from the map $R_4$,
\begin{equation}
R_4(\xi)=\frac{\xi^4+2\sqrt{3}i\xi^2+1}{\xi^4-2\sqrt{3}i\xi^2+1}  \ \ . \
\label{map2} \end{equation}
This rational map leads to the unit vector (\ref{unit2})
\begin{equation}
\vec {n}_{R_4}= \left(-\frac{(r^2-z^2)^2-2 z^2 r^2+2 x^2 y^2}
                        {{\cal N}_4} \ ,
                   \frac{\sqt (r^2+z^2)(x^2-y^2)}
                        {{\cal N}_4} \ ,
                   \frac{4 \sqt r x y z}
                        {{\cal N}_4}
             \right) \ ,
\label{unit4}
\end{equation}
with ${\cal N}_4 =2(x^4+x^2 y^2+x^2 z^2+y^4+y^2 z^2+z^4)$.

For static solutions with
$A_0=0$, $A_i = A_i(\vec{x}\,)$, $i=1,2,3$,
$\phi = \phi(\vec{x}\,)$,
the general set of field equations then involves the dilaton function
$\phi(\vec{x}\,)$, and nine functions for the gauge field, $A_i^a(\vec{x}\,)$.
The solutions are characterized by the rational map of degree $N$,
and, for a given map, numbered with increasing excitation by the integer $k$,
where the fundamental solution has $k=1$.

To obtain solutions with finite energy,
we demand that the gauge field of the fundamental
solution (and for odd $k$ solutions in general) 
tends to a pure gauge at infinity,
\begin{equation}
A_i = {i} (\partial_i U_\infty) U_\infty^\dagger
\ ,
\label{bcA}
\end{equation}
with $U_\infty=i \tau_R$, where
\begin{equation}
\tau_R := \vec {n}_R \cdot {\vec \tau}\ .
\label{bcHiggs}
\end{equation}
In contrast, the gauge field of the first excited solution
(and for even $k$ solutions in general) should vanish at infinity.

We also require that the dilaton field vanishes at infinity,
$\phi(\infty)=0$,
since any finite value of the dilaton field at infinity
can always be transformed to zero via
$\phi \rightarrow \phi - \phi(\infty)$,
when the radial coordinate is transformed according to
$r \rightarrow r e^{- \phi(\infty)} $.

The boundary conditions for platonic solutions are,
in principle, only needed at infinity. 
Specifying the unit vector $\vec {n}_R= \vec {n}_{R_4}$ thus completely
determines the boundary conditions for the gauge field 
of the fundamental $N=4$ platonic solution with cubic symmetry (\ref{bcA}).
The gauge field then exhibits at infinity
the angle-dependence of the rational map, 
since $U_\infty=i\tau_{R_4}$.
In contrast, for the first excited $N=4$ platonic solution
the discrete symmetry
does not enter via the boundary conditions at infinity,
since the gauge field vanishes there.

Parametrizing the gauge potential $A_i^a$ in terms of polar components,
$A_r^a$, $A_\theta^a$, and $A_\varphi^a$, we obtain a smooth numerical
solution for the first excited platonic $N=4$ YMD solution.
For the corresponding fundamental YMD solution, however,
this parametrization does not lead to an everywhere smooth numerical solution 
(with the same numerical method \cite{fidisol}).
The polar gauge field functions of the fundamental solution exhibit
a discontinuity at infinity, 
which we relate to the fact that 
the polar gauge field functions of the fundamental solution
are angle-dependent at infinity,
whereas those of the first excited solution approach zero at infinity.

Assuming that the observed discontinuity is a numerical effect, 
we expect to be able to avoid it by employing a better suited
parametrization for the gauge field.
In particular, the angular dependence of the gauge field at infinity
should be contained explicitly in the new parametrization
for the fundamental solution \cite{note1}.

To obtain such a parametrization, we note, that at infinity
$A_\theta$ and $A_\varphi$ approach
\begin{equation}
i \ (\partial_\theta U_\infty) U_\infty^\dagger = 
\frac{\partial_\varphi \tau_R}{\sin\theta} \ , 
\ \ \ \ {\rm and} \ \ \ \ \
i \ (\partial_\varphi U_\infty) U_\infty^\dagger = 
-\sin\theta \ \partial_\theta \tau_R \ 
\ , \end{equation}  
respectively,
since
\begin{equation}
 \vec {n}_R \times \partial_\theta \, \vec {n}_R =
\partial_\varphi \,\vec {n}_R \,/ \sin\theta \,
\ . \end{equation}
We therefore choose $\tau_R$, $\partial_\theta \tau_R$, 
and $\partial_\varphi \tau_R$ as a
basis of the Lie algebra of SU(2). 
This is adequate, as long as
$\partial_\theta \, \vec {n}_R$ and $\partial_\varphi \,\vec {n}_R$ do
not vanish. 
Since
for the rational map $R_4$, the vectors $\partial_\theta \, \vec {n}_{R_4}$ and 
$\partial_\varphi \,\vec {n}_{R_4}$ vanish along the cartesian coordinate axes,
problems may arise in the numerical proceduce, unless boundary
conditions are specified here.
Thus we demand that the gauge field vanishes along the coordinate axes.

Accordingly, we employ the following parametrization for the gauge field
of the fundamental solution,
\begin{equation}
A_\theta = f_\theta^1 \, \frac{\partial_\varphi \tau_R}{\sin\theta}
          +f_\theta^2 \, \frac{\partial_\theta \tau_R}{|\partial_\theta \vec {n}_R|}           
          +f_\theta^3 \, \tau_R  \ ,
\label{pA1}
\end{equation}
\begin{equation}
A_\varphi =   f_\varphi^1 \, (-\sin\theta \ \partial_\theta \tau_R) 
          +f_\varphi^2 \, \frac{\partial_\varphi \tau_R}{|\partial_\varphi \vec {n}_R|}     
          +f_\varphi^3 \, \tau_R  \ ,
\label{pA2}
\end{equation}
and
\begin{equation}
A_r =  f_r^1 \, \tau_x
      +f_r^2 \, \tau_y           
      +f_r^3 \, \tau_z  \ .
\label{pA3}
\end{equation}
The functions $f_i^a$ now have no angular dependence at infinity.
In particular,
$f_\theta^1$ and $f_\varphi^1$ approach unity at infinity,
while the other seven functions vanish there. 

Subject to the boundary conditions and to the gauge condition
\begin{equation}
\partial_i A^i =0 \ , 
\label{gaugecond} 
\end{equation} 
the set of field equations can be solved numerically.  The discrete
symmetries of the platonic solution allow for a restriction of the
region of numerical integration.  Specifying the reflection symmetries
of the fields with respect to the $xy$-, $xz$- and $yz$-plane, it is
sufficient to solve the field equations for $x\ge 0$ and $y\ge 0$ and
$z\ge 0$ only \cite{kkm-dil}. The integration region can even be
restricted to $x \ge y \ge 0$ and $z \ge 0$, with the reflection symmetries 
of the fields at $\varphi=\pi/4$ .

The boundary conditions at infinity must now be supplemented
by conditions at the other boundaries of the integration region.
These boundaries consist of the origin, parts of the $xy$-, $xz$- and $yz$-plane 
(or the plane $\varphi=\pi/4$),
and the positive $z$-axis.
(Together with the boundary at infinity these boundaries
are implemented numerically as the boundaries of a cube \cite{fidisol}.)
The gauge field vanishes at the origin,
and the dilaton field satisfies $\partial_{{r}} \phi = 0$.

The boundary conditions in the $xz$-plane ($\vphi=0$), the $yz$-plane
($\vphi=\pi/2$), in the plane $\vphi=\pi/4$, on the $z$-axis ($\theta=0$) and in the $xy$-plane
($\theta=\pi/2$) follow from the reflection symmetries of the fields.
In particular, we suppose that the rational map $R_4$ determines the
reflection symmetries of the fields.  Thus we impose, that for the
platonic YMD solutions the gauge field $A_i$ possess the same
reflection symmetries as ${i} [\partial_i \tau_{R_4}
,\tau_{R_4}]$.  For the gauge field
functions $A_i^a$ and $f_i^a$, the proper set
of boundary conditions is given in Table 1. For the dilaton field the normal derivative must
vanish in the $xy$-, $xz$- and $yz$-plane, in the plane $\varphi=\pi/4$, and on the $z$-axis
($\partial_\theta \phi =0$).

\begin{table}[t!]
\begin{tabular}{|c|c|c|}
\hline
$k$ & $xz$- and $yz$-plane ($\varphi=0$ or $\pi/2$) & $\varphi=\pi/4$
\\ \hline
1 & \begin{tabular}{ccc}
       $f_r^1=0$ \ , &   $f_r^2=0$ \ ,  &  $\partial_\vphi f_r^3=0$ \ , \\
       $\partial_\vphi f_\theta^1=0$ \ , &   $f_\theta^2=0$ \ ,  &   
       $f_\theta^3=0$ \ , \\
       $\partial_\vphi f_\vphi^1=0$ \ ,  &  $f_\vphi^2=0$ \ , &   
       $\partial_\vphi f_\vphi^3=0$ \ ,
     \end{tabular} 
   & \begin{tabular}{ccc}
       $f_r^1=0$ \ ,  &   $\partial_\vphi f_r^2=0 $\ ,   &   $f_r^3=0 $\ , \\
       $\partial_\vphi f_\theta^1=0$ \ ,   &   $f_\theta^2=0$ \ ,   &  
       $f_\theta^3=0$ \ , \\
       $\partial_\vphi f_\vphi^1=0$ \ ,   &   $f_\vphi^2=0$ \ ,   &  
       $\partial_\vphi f_\vphi^3=0$ \ ,
     \end{tabular} 
 \\ \hline
2 & \begin{tabular}{ccc}
       $A_r^1=0$ \ , &   $A_r^2=0$ \ ,  &  $\partial_\vphi A_r^3=0$ \ , \\
       $A_\theta^1=0$ \ , &   $A_\theta^2=0$ \ ,  &   
       $\partial_\vphi A_\theta^3=0$ \ , \\
       $\partial_\vphi A_\vphi^1=0$ \ ,  &  $\partial_\vphi A_\vphi^2=0$ \ , &   
       $A_\vphi^3=0$ \ ,
     \end{tabular} 
   & \begin{tabular}{ccc}
       $A_r^1=0$ \ ,  &   $\partial_\vphi A_r^2=0 $\ ,   &   $A_r^3=0 $\ , \\
       $A_\theta^1=0$ \ ,   &   $\partial_\vphi  A_\theta^2=0$ \ ,   &  $A_\theta^3=0$ \ , \\
       $\partial_\vphi A_\vphi^1=0$ \ ,   &   $A_\vphi^2=0$ \ ,   &  
       $\partial_\vphi A_\vphi^3=0$ \ ,
     \end{tabular} 
 \\ \hline
    & $z$-axis ($\theta=0$) & $xy$-plane ($\theta=\pi/2$) 
\\ \hline  
1 & \begin{tabular}{ccc}
     $f_r^1=0$ \ , & $f_r^2=0$ \ , & $\partial_\theta f_r^3=0$ \ , \\
     $\partial_\theta f_\theta^1=0$ \ , & $f_\theta^2=0$ \ , & $f_\theta^3=0$ \ , \\
     $\partial_\theta f_\vphi^1=0$ \ , & $f_\vphi^2=0$ \ , & $f_\vphi^3=0$ \ ,
     \end{tabular} 
  & \begin{tabular}{ccc}
     $f_r^1=0$ \ , & $f_r^2=0$ \ , & $\partial_\theta f_r^3=0$ \ , \\
     $\partial_\theta f_\theta^1=0$ \ , & $f_\theta^2=0$ \ , & 
     $\partial_\theta f_\theta^3=0$ \ , \\
     $\partial_\theta f_\vphi^1=0$ \ , & $f_\vphi^2=0$ \ , & $f_\vphi^3=0$ \ ,
     \end{tabular} 
 \\ \hline
2 & \begin{tabular}{ccc}
     $A_r^1=0$ \ , & $A_r^2=0$ \ , & $\partial_\theta A_r^3=0$ \ , \\
     $A_\theta^1=0$ \ , & $A_\theta^2=0$ \ , & $A_\theta^3=0$ \ , \\
     $A_\vphi^1=0$ \ , & $A_\vphi^2=0$ \ , & $A_\vphi^3=0$ \ ,
     \end{tabular} 
  & \begin{tabular}{ccc}
     $A_r^1=0$ \ , & $A_r^2=0$ \ , & $\partial_\theta A_r^3=0$ \ , \\
     $\partial_\theta A_\theta^1=0$ \ , & $\partial_\theta A_\theta^2=0$ \ , & 
     $A_\theta^3=0$ \ , \\
     $A_\vphi^1=0$ \ , & $A_\vphi^2=0$ \ , & $\partial_\theta A_\vphi^3=0$ \ .
     \end{tabular} 
 \\ \hline
\end{tabular} 
\caption{ \small
The boundary conditions 
in the $xz-$, $yz-$, and $xy$-plane, on the $z$-axis, and for $\varphi=\pi/4$
are given for the gauge field components
of the fundamental ($k=1$) and 
first excited ($k=2$) platonic solution.}
\end{table}

An adequate initial guess is mandatory for the numerical 
construction of platonic solutions.
The initial guess is obtained
by parametrizing the gauge field according to
\begin{equation}
A_i =
\frac{1-f(r)}{2} i\, (\partial_i \tau_{R_4} )\tau_{R_4}
\ , \label{guess1}
\end{equation}
and the dilaton field by $\phi(r)$. 
This simple ansatz with two radial functions
$f(r)$ and $\phi(r)$
is inserted into the action,
the angular dependence is integrated,
and a set of variational equations is obtained,
similar to the set of spherically symmetric ($N=1$) equations.
These ordinary differential equations are solved numerically,
subject to the set of boundary conditions of the spherical ($N=1$) solutions
(in particular $f(0)=1$, $f(\infty)=-1$ for the fundamental solution and 
$f(0)=1$, $f(\infty)=1$ for the first excited solution).
The initial guess is then given by (\ref{guess1})
with the numerically determined functions $f(r)$ and $\phi(r)$.

\section{\bf Properties of Platonic Solutions}

The numerical solutions are constructed with help of the 
software package FIDISOL \cite{fidisol} based on the Newton-Raphson 
algorithm. 
The solutions are obtained in spherical coordinates $r$, $\theta$, $\vphi$.
To map the infinite range of the radial variable $r$ to the finite 
interval $[0,1]$ we introduce the compactified variable $\bar{r}=r/(1+r)$.
Typical grids contain $70 \times 25 \times 25$ points.
The estimated relative errors are on the order of $1\% $, or better.

For static finite energy solutions (with $A_0=0$) 
the energy is given by
\begin{equation}
E = \int\left( \frac{1}{2} \partial_i \phi \partial^i \phi
+\frac{1}{2} e^{2 \phi} {\rm Tr}\, (F_{ij} F^{ij})
\right) \rd^3 x \ .
\label{energy}
\end{equation}
It is related to the dilaton charge $D$,
\begin{equation}
D = \frac{1}{4\pi} \int_{S_2} \vec{\nabla}\phi \cdot d\vec{S} \ ,
\end{equation}
by \cite{kkm-dil}
\begin{equation}
 D =  E \ .
\end{equation}
The energy density $\epsilon (\vec x)$ is defined by
\begin{equation}
E= \int \epsilon (\vec x) \rd^3 x
\ . \end{equation}

Let us begin the discussion of the
properties of the platonic YMD solutions
by briefly recalling the properties of
the spherically and axially symmetric
EYM and YMD solutions \cite{bm,ymd,kk,kk-dil}.
In both cases for an appropriate map of degree $N$,
an infinite sequence of solutions exists,
labelled by the node number $k$ of the gauge field function(s).

The energy $E$ of the solutions of the sequences 
corresponding to maps with $N=1-4$ is
shown in Table 2 \cite{kk,kk-dil},
and exhibits close agreement between the sets of EYM and YMD solutions.
\begin{table}[t!]
\begin{center}
\begin{tabular}{|c|cccc|cccc|}
 \hline
\multicolumn{1}{|c|} { $ $ }&
\multicolumn{4}{ c|}  {$EYM$ } &
\multicolumn{4}{ c|}  {$YMD$ } \\
 \hline
$k/N$ &  $1$  & $2$ & $3$ & $4$ 
      &  $1$  & $2$ & $3$ & $4$ \\
 \hline
$1$ &  $.829$   & $1.385$  & $1.870$  & $2.319$  
 &  $.804$   & $1.336$  & $1.800$  & $2.231$  \\
$2$ &  $.971$   & $1.796$  & $2.527$  & $3.197$  
 &  $.966$   & $1.773$  & $2.482$  & $3.137$  \\
$3$ &  $.995$   & $1.935$  & $2.805$  & $3.621$  
 &  $.994$   & $1.927$  & $2.785$  & $3.588$  \\
$4$ &  $.999$   & $1.980$  & $2.921$  & $3.824$  
 &  $.999$   & $1.977$  & $2.913$  & $3.808$  \\
\hline
$\infty$ &  $1.$   & $2.$  & $3.$  & $4.$  
 &  $1.$   & $2.$  & $3.$  & $4.$  \\
\hline
\end{tabular}
\end{center}
{\bf Table 2}\\
The energy $E$ of EYM solutions 
and YMD solutions 
with spherical symmetry ($N=1$)
and axial symmetry ($N=2-4$)
and node numbers $k=1-4$.
For comparison, the energy $E_\infty$ of the limiting
solutions of the sequences is also shown.
Note, that $E/N$ decreases with $N$ for fixed finite $k$.
\end{table}

In the limit $k \rightarrow \infty$,
the sequences of solutions corresponding to a certain map
of degree $N$ tend to limiting solutions.
For the EYM sequence, the limiting solution
corresponds to the exterior of an
extremal Reissner-Nordstr\"om (RN) solution with mass $M=N$,
and magnetic charge $P=N$.
In isotropic coordinates, the horizon of the extremal RN
is located at $r_{\rm H}=0$, and its metric reads
\begin{equation}
ds^2= -\left( \frac{r}{N+r} \right)^2 dt^2
 + \left( \frac{N+r}{r} \right)^2
   \left ( d r^2 + r^2 \left( d \theta^2
           +  \sin^2 \theta d\varphi^2 \right) \right)
\ . \label{rn1} \end{equation}
This metric is of the type of the metric Eq.~(\ref{metric1}) 
with 
\begin{equation}
e^{2 \phi} = \left( \frac{r}{N+r} \right)^2
\ , \label{rel1} \end{equation}
and a euclidian 3-dimensional metric.
Not surprisingly, for the YMD sequence the limiting solution 
has dilaton field \cite{ymd,kk-dil}
\begin{equation}
\phi_\infty = - {\rm ln} \left( 1 + \frac{N}{r} \right)
\ , \label{rel2} \end{equation}
clearly representing the same extremal RN solution.
Thus with increasing excitation EYM and YMD solutions
differ less and less, and converge to the same limiting solution.
Considering a YMD solution as an approximation to an EYM solution,
the approximation thus improves with increasing $k$
for fixed $N$.

\begin{figure}[t!]
\parbox{\textwidth}{
\centerline{
\mbox{\epsfysize=6.0cm \epsffile{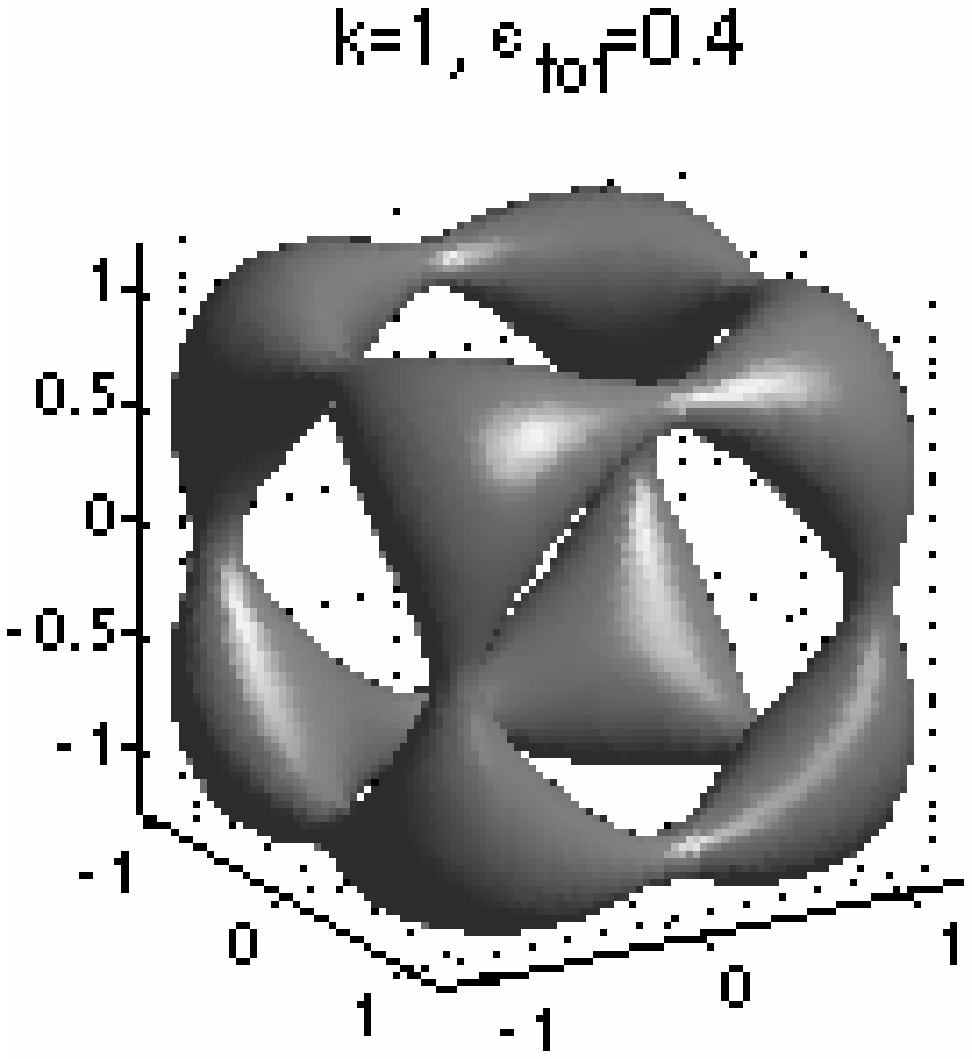} } \hspace{1cm}
\mbox{\epsfysize=6.0cm \epsffile{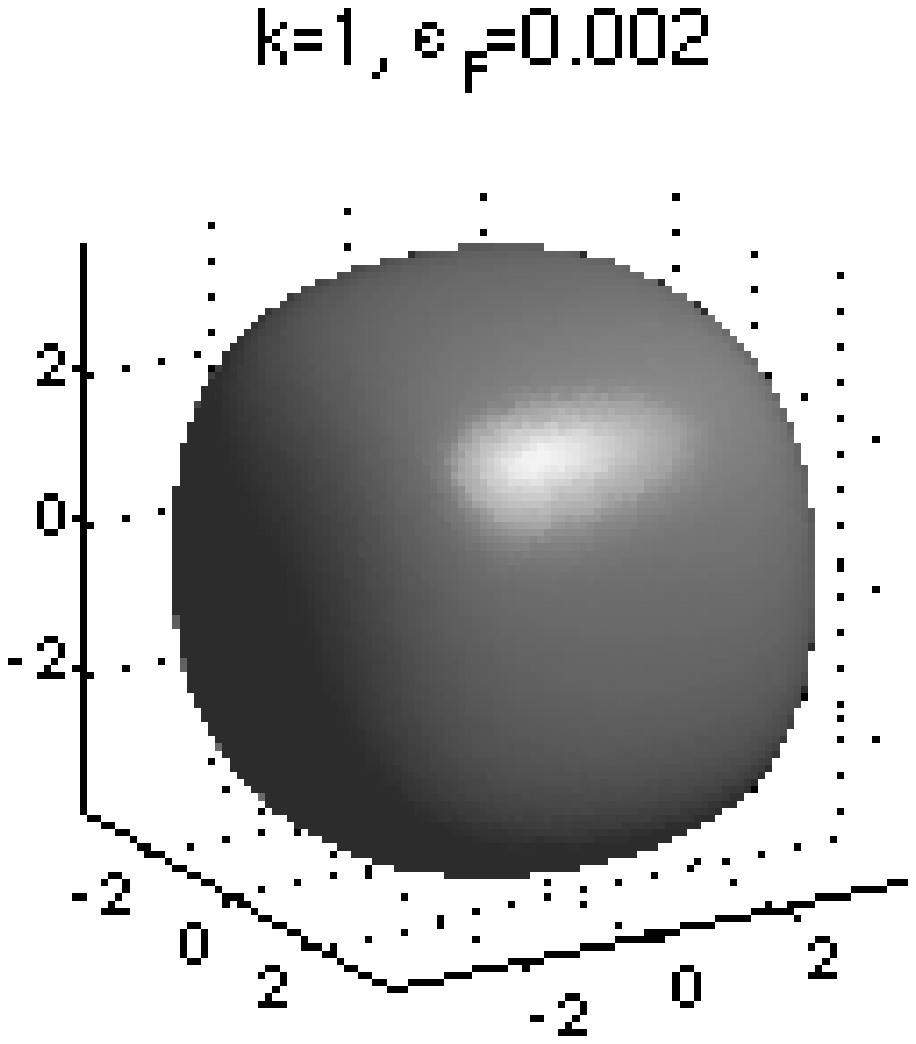} } 
}\vspace{1.cm} }
\parbox{\textwidth}{
\centerline{
\mbox{\epsfysize=6.0cm \epsffile{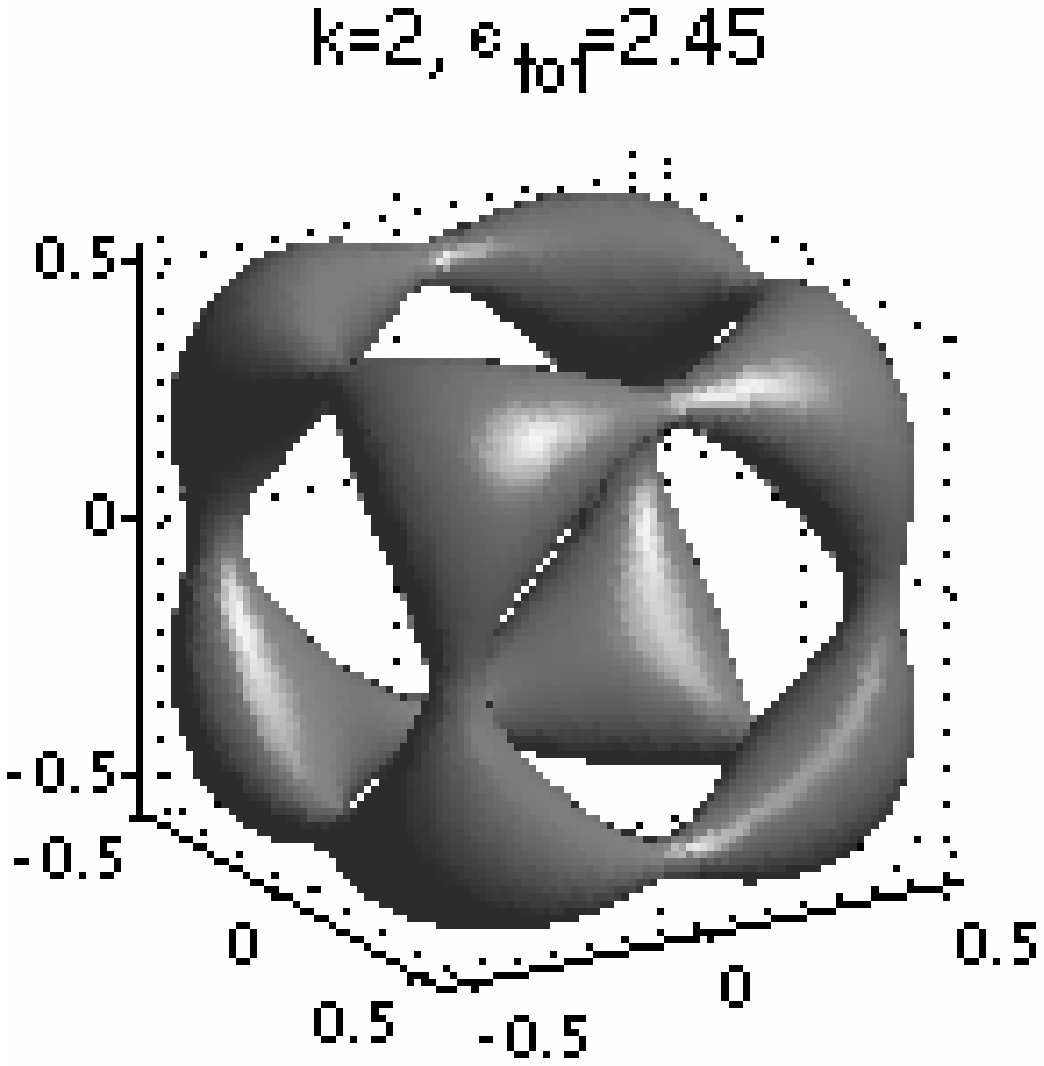} } \hspace{1cm}
\mbox{\epsfysize=6.0cm \epsffile{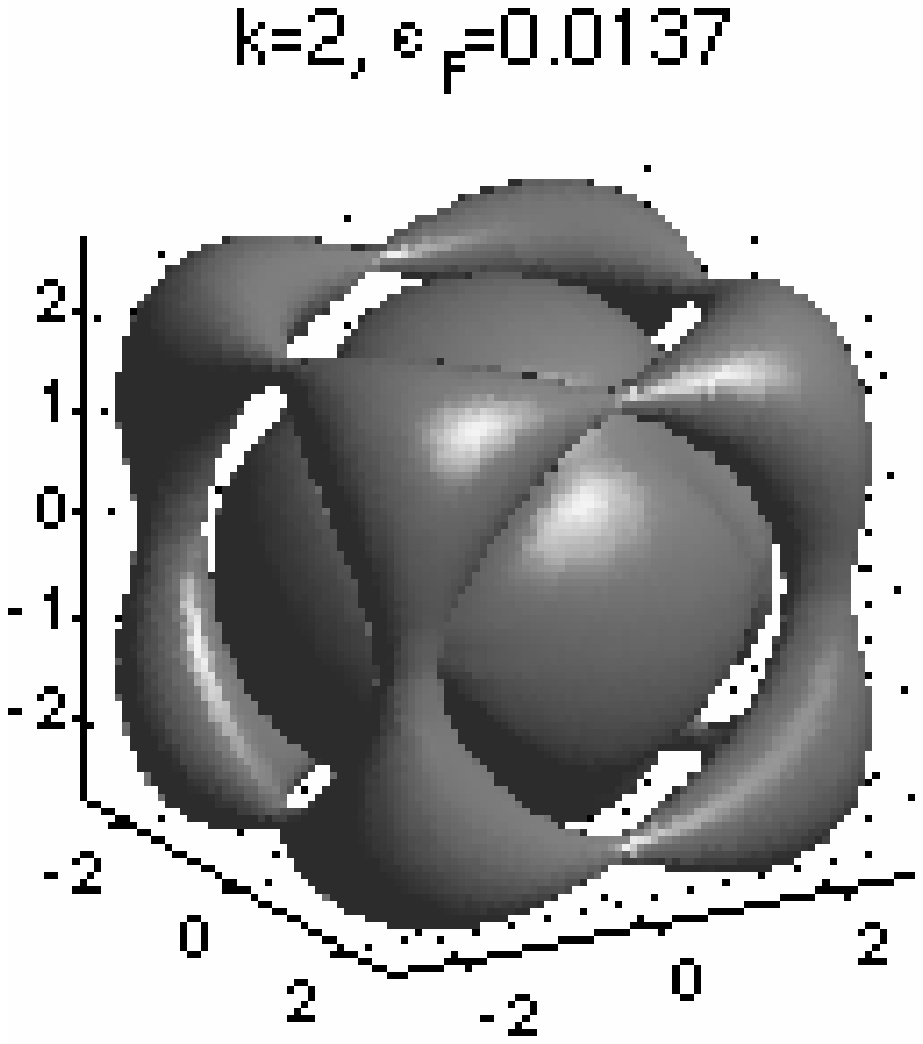} }
}\vspace{1.cm} }
{\bf Fig.~1} \small
Surfaces of constant total energy density $\epsilon_{\rm tot}$
(left column)
and of constant energy density of the gauge field $\epsilon_{\rm F}$
(right column)
are shown for the fundamental cubic YMD solution (upper row) and for the
first excitation (lower row).
\vspace{0.5cm}
\end{figure}
Let us now turn to the platonic YMD solutions.
In Fig.~1 
we present surfaces of constant total energy density $\epsilon_{\rm tot}$ 
and of constant energy density of the gauge field
$\epsilon_{\rm F}=
\frac{1}{2} e^{2\phi} {\rm Tr} (F_{ij} F^{ij})$
for the fundamental and first excited cubic solutions.
The energy density clearly reflects the symmetries of a cube.
As before \cite{sphapla,kkm-dil} we conclude
that the shape of the energy density of platonic solutions
is determined primarily by the rational map,
since we observe the same shape for the YMD solutions 
as for Skyrmions, sphalerons, and monopoles, based on the same map $R_4$
\cite{skyrpla,sphapla,monopla}.
The energy density of
the excited solution exhibits a cubic shape both for
large and for small constant values,
revealing a small interior cube within a larger exterior cube.

The energy of the fundamental cubic solution is $E_{k=1}=2.203$,
while the energy of the first excited cubic solution is $E_{k=2}=3.11$, 
with estimated errors on the order of $10^{-3}$ and $10^{-2}$, respectively.
Comparison of these values with those of the axially symmetric
$N=4$ solutions (Table 2) reveals, that the platonic solutions
have lower energy.\footnote{
Note that the difference in energy for the first excited solution is the
same order of magnitude as the estimated error though.}
For the cubic EYM solutions we conjecture that they possess
a slightly higher mass,
in accordance with the known rotationally symmetric solutions.
The observed values $E_{k=1}$ and $E_{k=2}$
are in agreement with the conjecture, that 
the energy of the excited cubic solutions tends to
the limiting value $E_\infty=4$ for $k\rightarrow \infty$.

The metric function $-g_{00}= e^{2 \phi}$
is exhibited in Fig.~2 for the fundamental and the first excited solution.
Clearly, the metric exhibits the same platonic symmetry
as the energy density.
\begin{figure}[t!]
\parbox{\textwidth}{
\centerline{
\mbox{\epsfysize=6.0cm \epsffile{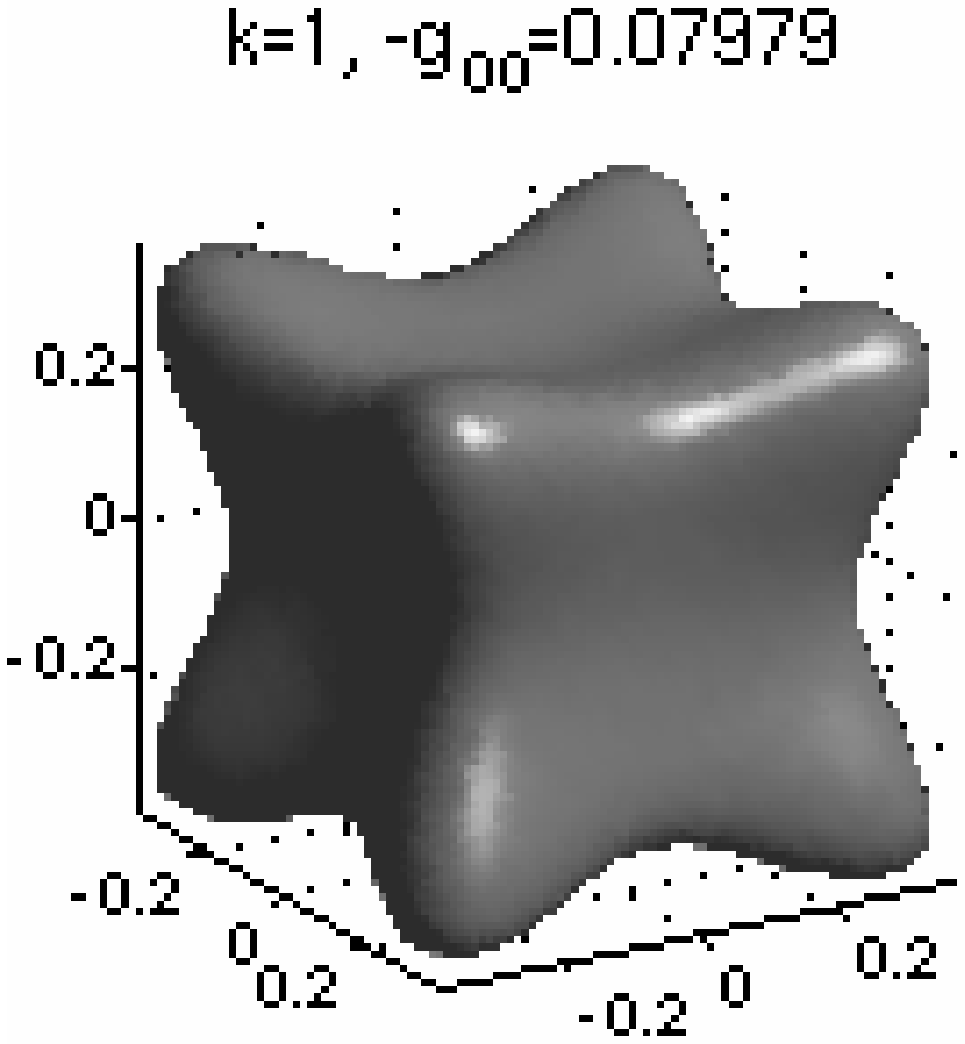} } \hspace{1cm}
\mbox{\epsfysize=6.0cm \epsffile{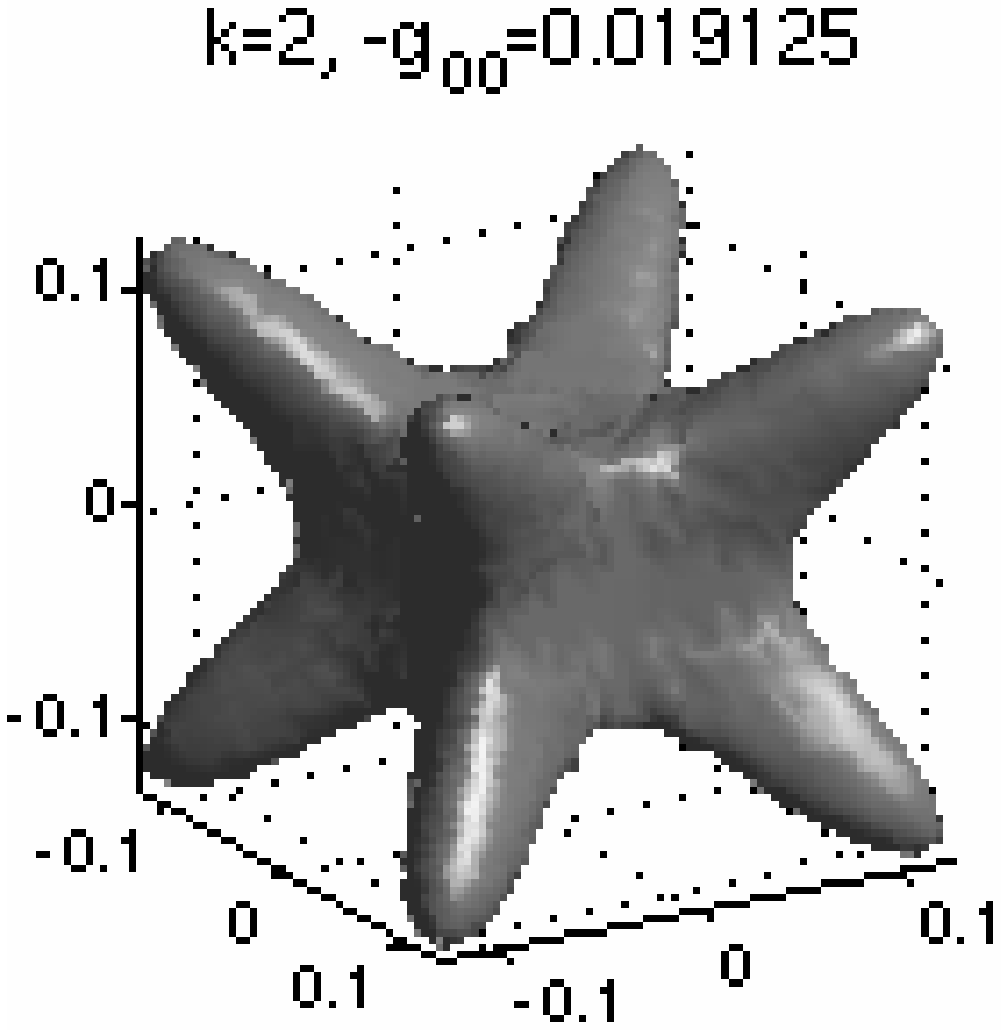} } 
}\vspace{1.cm} }
\parbox{\textwidth}{
\centerline{
\mbox{\epsfysize=6.0cm \epsffile{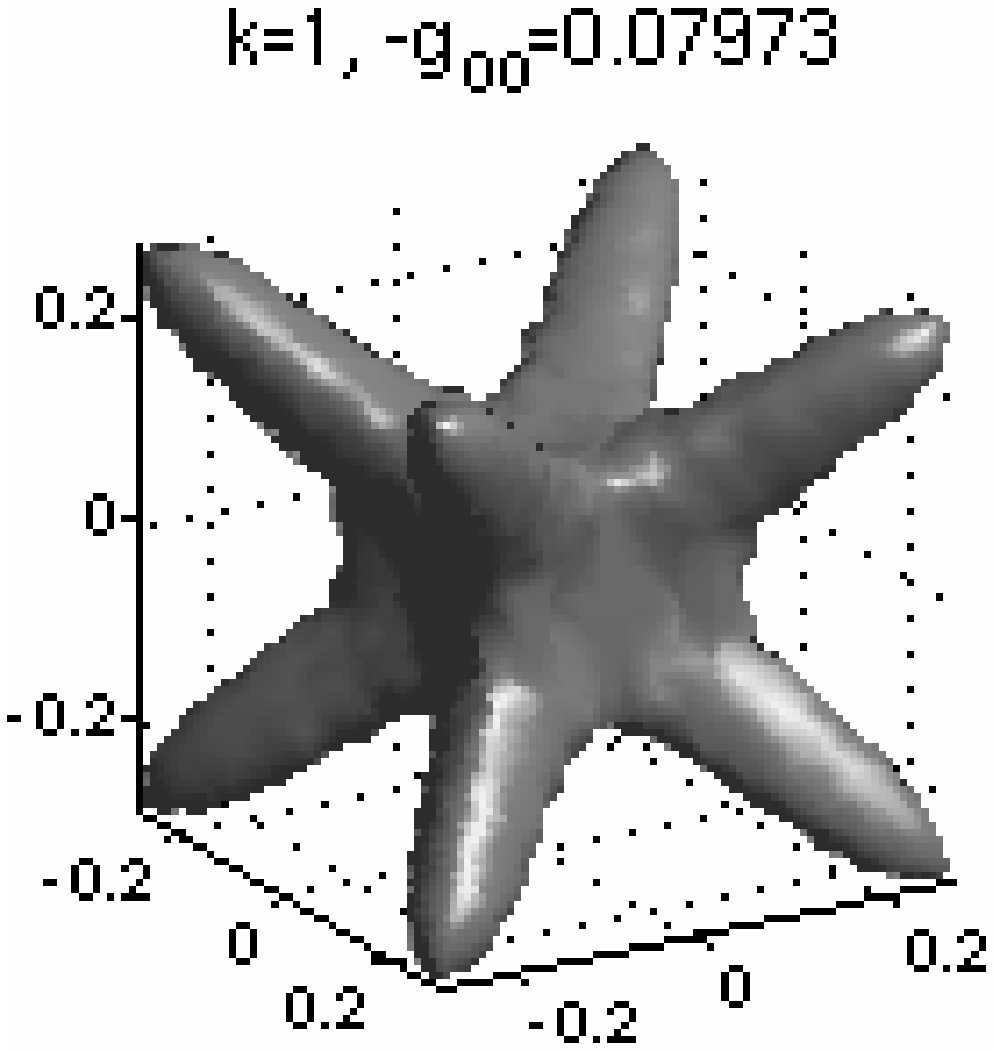} } \hspace{1cm}
\mbox{\epsfysize=6.0cm \epsffile{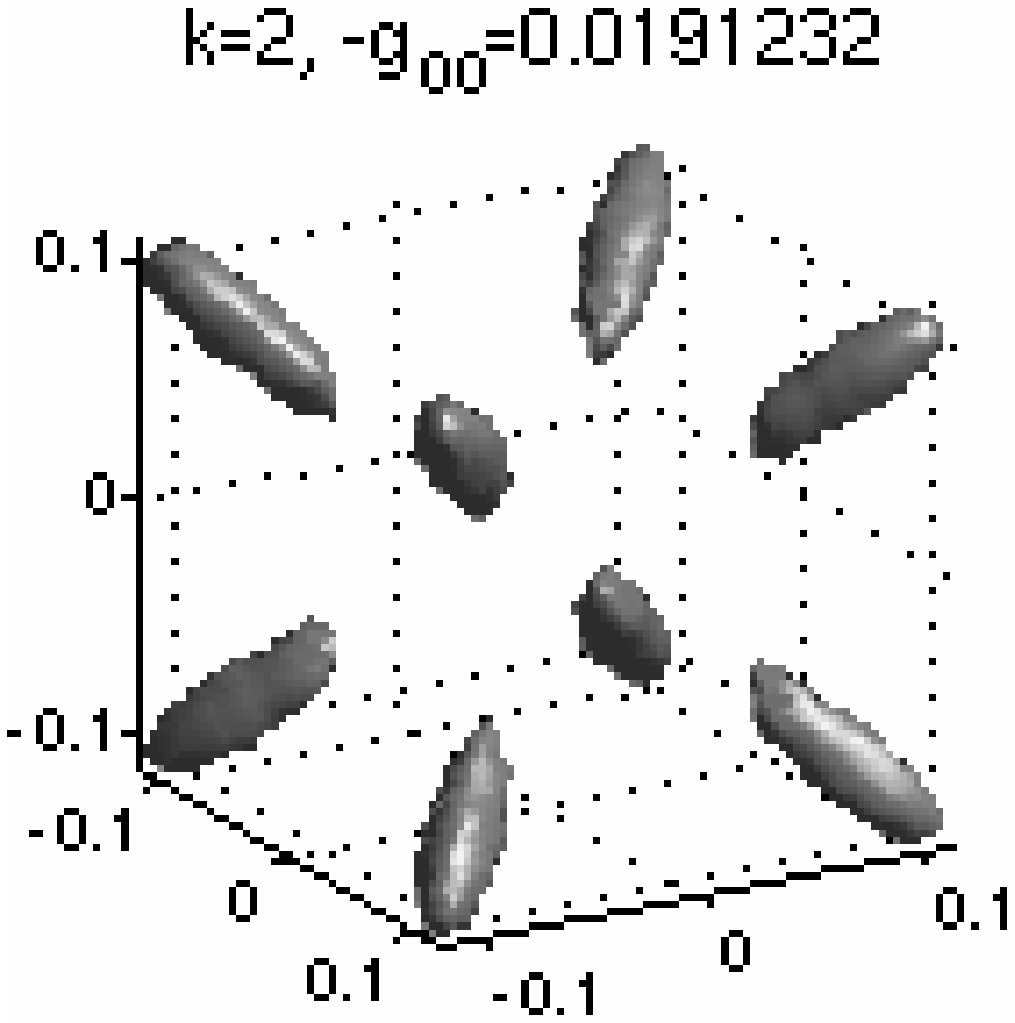} }
}}
{\bf Fig.~2} \small
Surfaces of constant metric function $-g_{00}= e^{2 \phi}$
for the fundamental cubic YMD solution (left column)
and the first excitation (right column).
\vspace{0.5cm}
\end{figure}

In Fig.~3 we further address the conjecture, that platonic solutions
associated with maps of degree $N$
form infinite sequences, tending to
extremal RN solutions with magnetic charge $P=N$.
The cubic solutions should therefore tend
to a limiting extremal RN solution with charge $P=4$.

We first consider convergence of the dilaton function $\phi_k(\vec x\,)$
to the limiting function $\phi_\infty(\vec x\,)$, Eq.~(\ref{rel2}). Therefore
we extract a `local magnetic charge' $P_k(\vec x\,)$ for the
fundamental ($k=1$) and first excited ($k=2$) solution, by defining
\begin{equation}
\phi_k(\vec x\,) = - {\rm ln} \left( 1 + \frac{P_k(\vec x\,)}{r} \right)
\ . \label{rel2a} \end{equation}
Exhibiting $P_k(\vec x\,)$ as a function of $r$ for three fixed angles,
we observe indeed that the functions $P_k(\vec x\,)$
approach the constant $P_\infty=4$ with increasing $k$ 
in an increasing interval.

We then extract a `local magnetic charge' $\tilde P_k(\vec x\,)$
from the metric component $g_{00}$ of a non-extremal
RN solution of mass $E$ in isotropic coordinates \cite{bm},
employing Eq.~(\ref{metric1})
\begin{equation}
\tilde P_k^2(\vec x\,) = 4 r^2 \left(
\frac{1 - e^{-\phi(\vec x\,)}}{1 + e^{-\phi(\vec x\,)}} + \frac{E^2}{4 r^2}
+ \frac{E}{r} \frac{1}{1 + e^{-\phi(\vec x\,)}} \right)
\ . \label{rel3} \end{equation}
Again we observe, that the charge $\tilde P_2(\vec x\,)$ obtained from
the first excited solution
is closer to the constant value $P_\infty=4$ of the
limiting extremal RN solution in a wider range
than the charge $\tilde P_1(\vec x\,)$ of the fundamental solution,
indicating that with increasing $k$ the platonic solutions 
indeed tend to the spherically symmetric extremal RN solution with $P=4$.
\begin{figure}[t!]
\parbox{\textwidth}{
\centerline{
\mbox{\epsfysize=6.0cm \epsffile{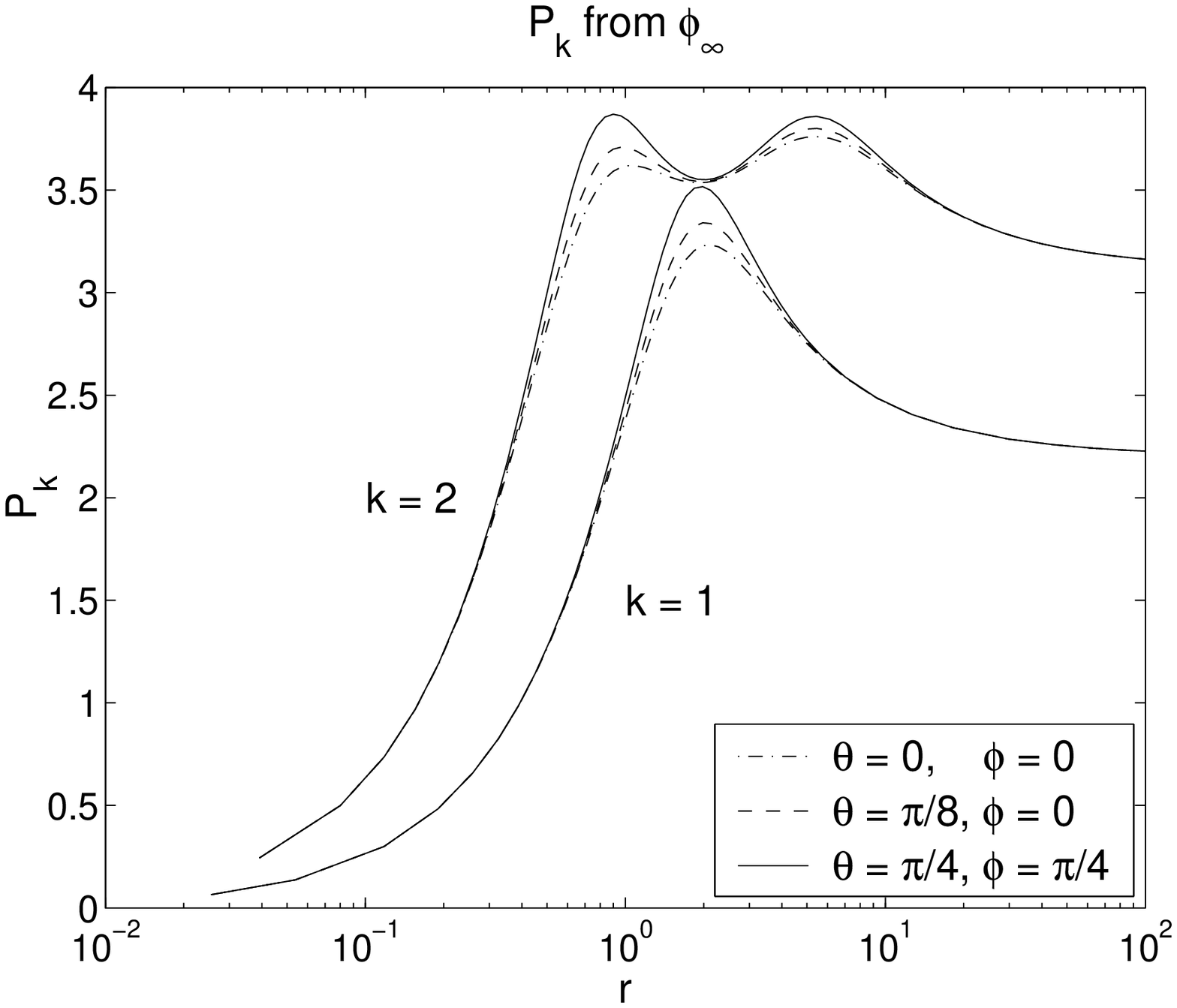} }
\mbox{\epsfysize=6.0cm \epsffile{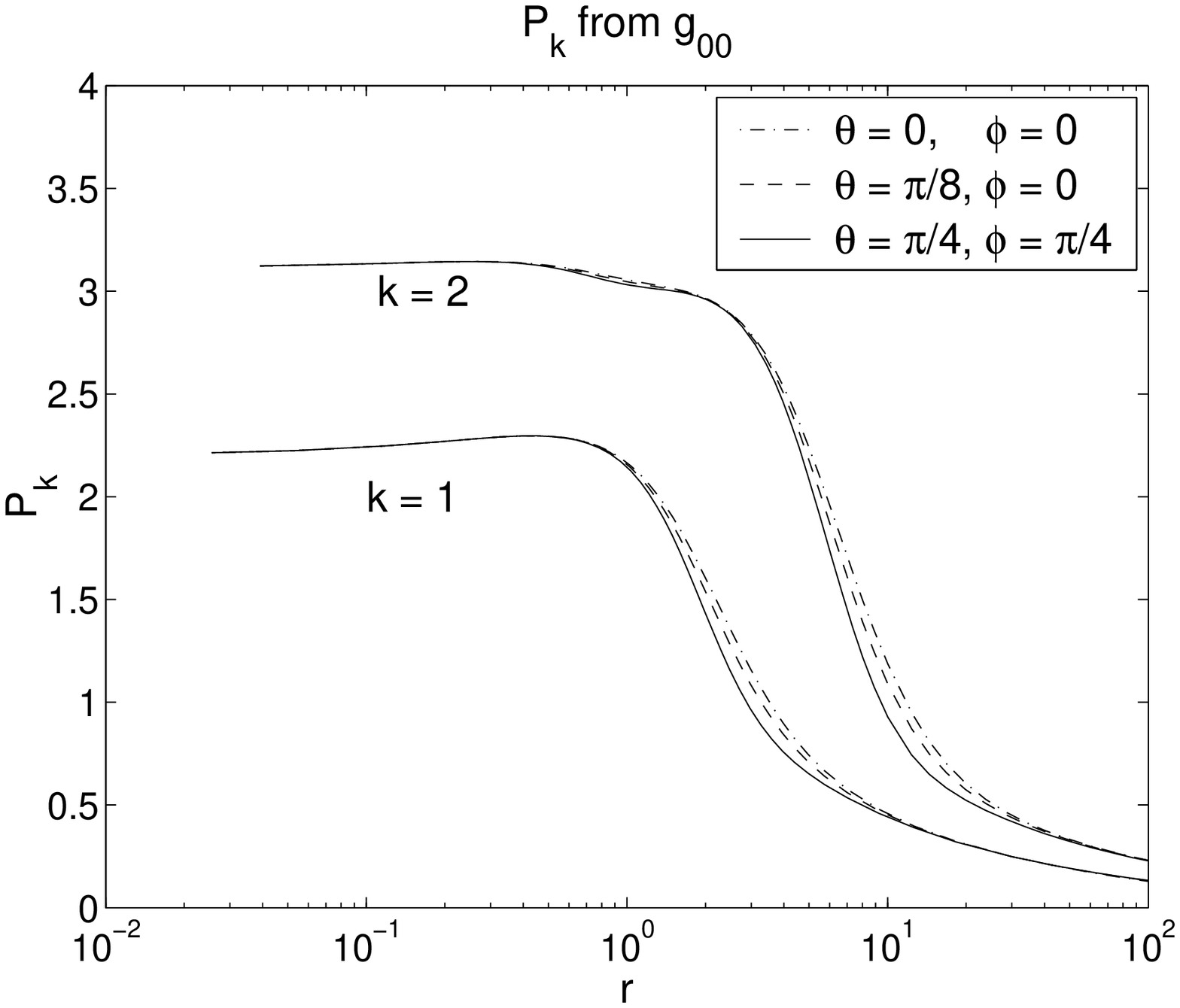} }
}\vspace{0.5cm}
{\bf Fig.~3} \small
The `local magnetic charge' $P_k(\vec x\,)$
extracted from Eq.~(\ref{rel2a}) (left) and $\tilde P_k(\vec x\,)$
extracted from Eq.~(\ref{rel3}) (right)
for the fundamental ($k=1$) and first excited ($k=2$) cubic solution
for three representative angles.
\vspace{0.5cm}
}
\end{figure}

\section{Conclusions and Outlook}

Motivated by the aim to demonstrate the existence of
non-perturbative gravitating solutions with only discrete symmetries,
we have constructed platonic solutions in YMD theory.
These solutions 
may be viewed as exact (numerical) solutions of scalar gravity,
by considering the dilaton as a kind of scalar graviton,
or as approximate solutions of Einstein-Yang-Mills theory,
when the spatial part of the metric is taken to be euclidian.

We have focussed on platonic solutions with cubic symmetry,
related to a certain rational map of degree $N=4$.
By constructing numerically the fundamental cubic solution
and its first excitation,
we have here presented the first two solutions of the cubic $N=4$ sequence.
We conjecture, that the sequence converges to the extremal
Reissner-Nordstr\"om solution with magnetic charge $P=4$.

By including the dilaton to approximate the effects of gravity,
we have obtained evidence for the existence of platonic EYM solutions,
while avoiding the complexity of the full set of EYM equations
in the absence of rotational symmetry.
Construction of exact (numerical) platonic EYM solutions, however,
remains a true challenge.

The existence of gravitating regular solutions
involving non-Abelian fields, is related to the existence of
black holes with non-Abelian hair \cite{bh}.
Thus the existence of gravitating regular solutions with platonic symmetries
strongly indicates the existence of a completely new type
of black holes: static black holes which possess only
discrete symmetries \cite{ridgway}.

{\bf Acknowledgement}: 

B.K.~gratefully acknowledges support by the DFG under contract
KU612/9-1, and K.M.~by the Research Council of Norway under 
contract 153589/432.



\begin{thebibliography}{000}
\bibitem{note1}
 This problem did not occur for fundamental solutions
 of YMH and YMHD theory, due to the presence of the Higgs field
 \cite{sphapla,kkm-dil}.
\end{thebibliography}
\end{document}